\documentstyle[aps,12pt]{revtex}
\def\beginapjbib{\begingroup \section*{References}
         \parskip=.5ex plus 1.0pt
	 \def\bibitem{\par \noindent \hangindent\parindent
		\hangafter=1}}
\def\endapjbib{\par \endgroup}

\def\endapjbib{\par \endgroup}
\begin{document}
\def\ApJ{{\sl Astrophys. J.}}
\def\la{\mathrel{\mathpalette\fun <}}
\def\ga{\mathrel{\mathpalette\fun >}}
\def\fun#1#2{\lower3.6pt\vbox{\baselineskip0pt\lineskip.9pt
  \ialign{$\mathsurround=0pt#1\hfil##\hfil$\crcr#2\crcr\sim\crcr}}}
\thispagestyle{empty}
\begin{center}
\vspace{0.3in}
{\bf On the Origin of the Clustered QSO Metal Absorption Lines}
{\ \\\ \\Xiangdong Shi\\
{\it Department of Physics, Queen's University,\\
Kingston, Ontario, Canada, K7L 3N6\\
E-mail: shi@astro.queensu.ca\  \  \  WWW: http: //astro.queensu.ca/~shi/}}
\vskip 1.2cm
\end{center}
\centerline{Abstract}
\bigskip

\bigskip

Observations show that there is significant clustering of
QSO metal absorption lines within the range of velocity dispersion between
200km/sec and 600 km/sec.
With a reasonable supernova rate, it is shown that
high velocity gases driven by SNe and/or strong stellar winds
could explain the clustered absorptions,
provided that QSO metal-line absorbers are
galactic halos or dwarf galaxies. Rich clusters of galaxies,
on the other hand, cannot yield the observed clustering of
QSO metal absorption lines.
\bigskip

\noindent {\it Subject Headings} : Galaxies: Quasars: absorption
lines---Stars: supernovae
\newpage
\pagestyle{plain}
\newpage

\section{Introduction}
The metal absorption lines in QSO spectra (type C in the classification
of Weyman, Carswell \& Smith (1981))
have raised great interests because
they might reveal properties and evolution of objects associated
with galaxies at redshifts $z\ga 1$.
The properties of these metal absorption lines
were reviewed by Weyman, Carswell \& Smith (1981) and recently by
York {\em et al.} (1991).
These absorption lines are commonly associated with
halos of galaxies (Bahcall \& Spitzer 1969), since halos contain
metals, have star formation, and are clustered like galaxies. It is
also reasonable if considering that the cross section of the absorbers
(if they associate with galaxies) needed to explain
these lines is $\sim 10$ times the visible extent of modern galaxies
(Sargent et al. 1979). Three options for ``halos'' have been proposed:
1. each absorber corresponds to a single halo (Pettini et al. 1983);
2. a small number of gas-rich dwarf galaxies
(like the Magellanic clouds) form each halo (York et al. 1986);
3. many mini-absorbers (with masses much less than
those of dwarf galaxies, such as 10$^3M_\odot$)
in a halo provide the metal-line absorption (Sargent et al. 1979).

A significant portion of these metal-line absorption systems are systems that
have multiple absorption lines with velocity splits $> 200$ km/sec, and
lie in the intermediate position between quasars and observers
(Pettini et al. 1983; York et al. 1984;
Sargent, Boksenberg, \& Steidel 1988, hereafter SBS).
A two-point correlation calculation of CIV absorption lines in QSO
spectra shows that there is significant correlation in the range of
200 km/sec $<\Delta V<600$ km/sec, where
$\Delta V$ is the velocity split between a pair of absorption lines (SBS).
In the survey of 55 QSOs by SBS, 111 CIV absorbers are
identified with an equivalent width larger than 0.15\AA\  and
a relative velocity with respect to
the QSO larger than 5000 km/sec; among them 18 pairs of
absorbers are found in excess of what would be expected from a random
Poisson distribution of absorbers in the range of 200 km/sec
$<\Delta V<600$ km/sec,
while no significant correlation is found in the range of
600 km/sec$<\Delta V\la 1000$ km/sec. In other words,
metal absorption lines are found to cluster in
the range of velocity dispersion between 200 km/sec and 600 km/sec.

The clustering of absorption lines in the range of 200 km/sec$<\Delta V
<600$ km/sec cannot, at least mainly, be explained by the
differential rotation motion within a galaxy, which typically yields
a $\Delta V$ of 150 km/sec. Attempts have been made to explain
these clustered absorptions by galaxy-galaxy correlation
at $z\sim 2$, the average redshift of these absorbers
(SBS). However, it was shown that the amplitude of today's galaxy-galaxy
correlation function extrapolated to $z\sim 2$, or
the galaxy-galaxy correlation calculated from cold dark
matter model, fell short to explain the amplitude of the clustering among
these absorbers by a factor $\ga$3 (SBS\footnote{the expected
correlation amplitude calculated by SBS from their eq.~(24) seems too high.
Using all the notations in SBS, it is clear from their eq.~(24) that
$n_{\rm c}(R)<2\pi R^2\int_{6R_\ast}^\infty\Phi (r_{\rm c}/r)^{1.8}dr$.
Since $\Phi =\Phi_0(1+z)^3$ and $r_{\rm c}=5h^{-1}(1+z)^{-5/3}{\rm Mpc}$ (SBS),
the number of correlated galaxies per galaxy divided by the mean number
of absorbers per unit redshift,
$n_{\rm c}/N<2.5\cdot 5h^{-1}{\rm Mpc}\cdot(5h^{-1}
{\rm Mpc}/6R_\ast)^{0.8}/(d_{\rm H}\sqrt
{1+z})$. With $R_\ast\approx 70h^{-1}$kpc, $d_{\rm H}=3000h^{-1}$Mpc
and $z\sim 2$, $n_{\rm c}/N<0.003$, a factor of 20 smaller than the 0.062
quoted in SBS. If a gravitationally unbound galaxy-galaxy correlation
is assumed, i.e., $r_0=5(1+z)^{-1}$Mpc, the resultant $n_{\rm c}/N
<0.01$, which still a factor of 6 lower than quoted in SBS.};
Salmon \& Hogan 1986). Although a high biasing factor may
improve the fit of the galaxy-galaxy correlation function to the
absorber correlation at $200$km/sec$<\Delta V<600$km/sec, it may be
hard to understand why a significant correlation of absorbers doesn't
extend to $\Delta V>600$ km/sec, given the high biasing and a
not-so-steep power law galaxy-galaxy correlation function (Peebles 1993)
\begin{equation}
\xi (r)\propto\Bigl({r_0\over r}\Bigr)^{-1.8},
\label{correl}
\end{equation}
where $r$ is the distance between a pair of galaxies and $r_0$ is a constant.

In this letter, we set aside the galaxy-galaxy correlation explanation
to these clustered absorption systems, and investigate
two other alternatives: 1) clusters of galaxies which have velocity
dispersions as high as $1000$ km/sec intercepting quasar lines of sight;
2) high velocity gases inside absorbers driven by
supernova shocks and/or stellar
winds. We will show that while the first alternative fails to explain
the observed clustering of QSO metal absorption lines at 200km/sec
$<\Delta V<600$km/sec, the second alternative may provide a viable
explanation.

\section{Clusters of galaxies}

If a quasar line of sight strikes two or more galaxies in a rich cluster,
they can provide clustered absorptions with a velocity spread as high as
$\sim 1000$ km/sec, which is the typical velocity dispersion of a
cluster (Pettini et al. 1983). However, since rich clusters only have
a dense core within a $0.2h^{-1}$ Mpc radius (Peebles 1993),
the chance of a quasar
line of sight striking multiple members of a single cluster is small.

Observations show that average rich clusters have a galaxy density
distribution of (Lilje \& Efstathiou 1988)
\begin{equation}
n(r)=n_{\rm b}[1+(r_0/r)^{-2.2}], \quad{\rm for\ }
0.2h^{-1}{\rm Mpc}\la r\la 20h^{-1}{\rm Mpc},
\label{c-g}
\end{equation}
where $n(r)$ is the number density of galaxies at a distance $r$ to
the cluster center, $n_{\rm b}$ is the background number density of
galaxies, $r_0\approx 9h^{-1}(1+z)^{-1.5}$Mpc, and
$h=H_0$/100km/sec/Mpc with $H_0$ being the Hubble constant.
 From APM survey $n_{\rm b}\approx 0.014 h^3(1+z)^3$
Mpc$^{-3}$  (Loveday et al. 1992). $n(r)$
falls off much faster beyond $20h^{-1}$Mpc (Lilje \& Efstathiou 1988),
and is much smoother inside the cluster core. In our following
calculations, it is convenient to assume a uniformly distributed core with
\begin{equation}
n(r)=n_{\rm core}\la (2R_{\rm h})^{-3},\quad{\rm for\ }r<0.2h^{-1}
{\rm Mpc},
\label{Ncore}
\end{equation}
where $R_{\rm h}$ is the typical radius of a galactic halo. The detailed
functional form of $n(r)$ in the core, however, does not affect the
generality of our calculations in any significant manner.
If each galactic halo provides each individual absorber, to explain the
number of
metal-line absorption systems observed, it was shown that (SBS)
\begin{equation}
R_{\rm h}\sim 70h^{-1} {\rm kpc}.
\label{Rh}
\end{equation}
Therefore, when a quasar line of sight passes
through the cluster with the closest distance $r_\perp$ from the cluster center
(as shown in Figure 1), the number of galaxies that the quasar
line of sight would intercept is
\begin{equation}
N(r_\perp)=\cases{2\pi\int_0^{\sqrt{R^2_{\rm c}-r^2_\perp}}
R^2_{\rm h}n(r)dr^\prime, &{\rm for}\  $r_\perp>r_{\rm c}$; \cr
2\pi\int_0^{\sqrt{r^2_{\rm c}-r^2_\perp}}
R^2_{\rm h}n_{\rm core}dr^\prime
+2\pi\int_{\sqrt{r^2_{\rm c}-r^2_\perp}}^{\sqrt{R^2_{\rm c}-r^2_\perp}}
R^2_{\rm h}n(r)dr^\prime,&{\rm for}\  $r_\perp<r_{\rm c}$.\cr}
\label{N}
\end{equation}
In eq.~(\ref{N}), $r=\sqrt{{r^\prime}^2+r^2_\perp}$. $R_{\rm c}$ is
the radius of the cluster. For the moment, we assume $R_{\rm c}$ to be
$20h^{-1}$ Mpc, the fall-off point in eq.~(\ref{c-g}). According to
Poisson statistics, the probability that this line of sight strikes
two or more galaxies in the cluster is
\begin{equation}
P(r_\perp)=1-[1+N(r_\perp)]e^{-N(r_\perp)}.
\label{P}
\end{equation}
Averaging over $r_\perp$, the probability that a quasar line of sight
strikes two or more galaxies of a cluster when it passes through the
cluster is $\int_0^{R_{\rm c}}P(r_\perp)\cdot 2r_\perp dr_\perp/R^2_{\rm c}$.
Considering that the number of
clusters the quasar line of sight passes through is
\begin{equation}
N_{\rm cl}=\int^{z_{\rm max}}_{z_{\rm min}}
\pi R_{\rm c}^2n_{\rm cl}d_{\rm H}(1+z)^{0.5}dz,
\label{Ncl}
\end{equation}
where $d_{\rm H}=3000h^{-1}$Mpc is the Hubble radius, and $n_{\rm cl}$ is
the number density of rich clusters today,
estimated to be $\approx 10^{-5}h^3$Mpc$^{-3}$ (Bahcall 1988).
The factor $(1+z)^{0.5}$ is the product of $(1+z)^3$ contributed by the number
density of rich clusters at $z$, $(1+z)^{-1.5}$ contributed by the
differential proper length, and $(1+z)^{-1}$ due to the scaling of
the proper length at $z$. A universe with the critical density is assumed.
The expected number of events for a quasar line of sight to strike
two or more galaxies in the same cluster, therefore, is
\begin{equation}
{\cal N}_{\rm multi}=2\pi\int^{z_{\rm max}}_{z_{\rm min}}
n_{\rm cl}d_{\rm H}(1+z)^{0.5}dz\int_0^{R_{\rm c}}P(r_\perp)\cdot
r_\perp dr_\perp.
\label{event}
\end{equation}
With a typical range of $z_{\rm min}\sim 1$ and $z_{\rm max}\sim 3$
for CIV absorption lines (SBS),
numerical integration of eq.~(\ref{event}) yields an event
rate of 0.06 per quasar. ${\cal N}_{\rm multi}$
would remain roughly unchanged even if we assume
eq.~(\ref{c-g}) is valid beyond 20 Mpc. It will only increase by
25$\%$ if we assume that the halo radii of
galaxies in the core of clusters are typically 35$h^{-1}$kpc instead of
eq.~(\ref{Rh}) (while the halo radius of galaxies outside the core remains
unchanged to maintain the required cross section to intercept quasar lines
of sight), resulting a $n_{\rm core}$ eight times higher according to
eq.~(\ref{Ncore}). Therefore, the expected number of pairs of absorbers
with 200km/sec$<\Delta V\la 1000$km/sec due to clusters of galaxies
intercepting QSO line of sight is $\la 0.06$ per quasar, or 3 for a
survey of 55 quasars, which can not explain
the $\sim 20$ pairs observed in this velocity range for 55 QSOs (SBS).

\section{High velocity gases in galaxies}
It is well known that supernova (SN) shocks or stellar winds can accelerate
the interstellar medium to velocities as high as $\sim 1000$km/sec.
Therefore, the clustered absorption systems in the range
200km/sec$<\Delta V<$600km/sec may originate from the activity of
massive stars inside the absorber itself, regardless of whether
the absorber is a single galaxy halo (Pettini et al. 1983),
a dwarf galaxy (York et al. 1986) or a mini-absorber
inside a galaxy (Sargent et al. 1979). A SN typically
releases 10$^{51}$ ergs in kinetic energy. A 30$M_\odot$ star also
deposits $\sim 10^{51}$ ergs in kinetic energy through mass loss during its
main sequence and pulsation life time of $\sim 5\times 10^6$ yr (Abbott 1982).
Simply on an energetic base,
if we assume negligible energy loss,
the amount of mass a SN shock or stellar wind can accelerate to a velocity of
$v$ for each SN or massive star is
\begin{equation}
M_v={10^{51}{\rm ergs}\over 0.5v^2}.
\label{Mv}
\end{equation}
In the phase in which SN shocks expand and decelerate to $\ga 10^2$km/sec,
the SN shocks do in fact propagate adiabatically with negligible energy loss
according to a power law (as given by the Sedov solution) (Spitzer 1978)
\begin{equation}
v\propto t^{-3/5},
\label{Powerlaw}
\end{equation}
where $t$ is the time of the propagation. In the case of stellar winds,
the gases pushed by the strong
stellar wind also propagate with negligible energy
loss at $v\ga 100$ km/sec with a power law (Weaver et al. 1977)
\begin{equation}
v\propto t^{-2/5}.
\end{equation}
Therefore, for $v\sim 100$km/sec, $M_v\sim 10^4M_\odot$.

We first deal with the case of SN shocks.
In our calculations, we assume that gases are uniformly
distributed in absorbers,
and assume that the absorber and the high velocity
region driven by SN shocks are spherical.
The geometry of a high velocity cloud due to a SN shock is illustrated
in Figure 2. When a quasar line of sight penetrates the cloud, a pair of
absorption lines are generated with a velocity dispersion
\begin{equation}
\Delta V=2v\cos\theta.
\label{DeltaV}
\end{equation}
To yield $\Delta V>200$km/sec, $v$ has to be larger than 100km/sec.
The average cross section of a high velocity cloud driven by a SN shock
to generate a $\Delta V>200$ km/sec absorption pair before the shock
decelerate to 100km/sec is

$$
\langle A\rangle_{\Delta V\ge 200{\rm km/sec}}
=\int_0^{t_{100{\rm km/sec}}}\pi r^2\Bigl[1-\Bigl({100{\rm km/sec}\over
v}\Bigr)^2\Bigr]{dt\over t_{100{\rm km/sec}}}\quad\quad\quad\quad
\nonumber$$

\begin{equation}
=\pi r^2_{\rm abs}\int_0^{t_{100{\rm km/sec}}}\Bigl({M_v\over M_{\rm abs}
}\Bigr)^{2/3}\Bigl[1-\Bigl({100{\rm km/sec}\over v}\Bigr)^2\Bigr]
{dt\over t_{100{\rm km/sec}}},
\label{<A>}
\end{equation}
where $r$ and $v$ are the radius and the velocity of the high velocity
gases respectively, $r_{\rm abs}$ is the radius
of the absorber in which the high velocity gases
is  embedded (which is either a single galaxy halo, or a dwarf galaxy, or a
mini-absorber inside a galaxy), and $M_{\rm abs}$ is the mass
of the absorber. $t_{100{\rm km/sec}}$ is the life time of the SN shock
when its velocity drops to 100km/sec.
After combining eq.~(\ref{Mv}), (\ref{Powerlaw}), (\ref{DeltaV})
and eq.~(\ref{<A>}), we get
\begin{equation}
\langle A\rangle_{\Delta V\ge 200{\rm km/sec}}
={2\over 9}\pi r^2_{\rm abs}\Bigl({M_{100{\rm km/sec}}
\over M_{\rm abs}}\Bigr)^{2/3},
\label{<A>2}
\end{equation}
and
\begin{equation}
t_{100{\rm km/sec}}={2\over 5}{r_{\rm abs}\over 100{\rm km/sec}}
\Bigl({M_{100{\rm km/sec}}\over M_{\rm abs}}\Bigr)^{1/3},
\label{t100}
\end{equation}
where $M_{100{\rm km/sec}}$ is $M_v$ at $v=100$km/sec.

Since the survey of SBS showed that $\sim 20$ pairs of CIV absorption lines
in the range of 200km/sec$<\Delta V<$600 km/sec were found in the total
of $\sim 100$ pairs of CIV absorption lines (within a sample of 111 CIV
absorption line systems), it implies that at any time the
cross section of these high velocity gases driven by SN shocks
to intercept a quasar line of light is
$\ga$10$\%$ of that of the absorbers. Therefore,
it can be estimated that
\begin{equation}
{R_{\rm SN}t_{100{\rm km/sec}}\langle A\rangle_{\Delta V\ge 200{\rm km/sec}}
\over\pi r^2_{\rm abs}}\ga 0.1,
\end{equation}
where $R_{\rm SN}$ is the SN rate in the absorber. The requirement on
$R_{\rm SN}$ is then
\begin{equation}
R_{\rm SN}>\Bigl({100{\rm km/sec}\over r_{\rm abs}}\Bigr)
\Bigl({M_{\rm abs}\over M_{100{\rm km/sec}}}\Bigr)\approx
\Bigl({100{\rm km/sec}\over r_{\rm abs}}\Bigr)
\Bigl({M_{\rm abs}\over 10^4M_\odot}\Bigr).
\label{SNrate}
\end{equation}

In the case of high velocity regions being driven by strong stellar winds,
the only difference from the SN case
is the power law dependence of $v$ on $t$, which only results in
a modification of eq.~(\ref{<A>2}) and (\ref{t100}) by a factor close
to 1. Since massive stars with strong stellar
winds typically have a mass $> 10M_\odot$, they inevitably end as
SNe. Therefore, our result, eq.~(\ref{SNrate}), will not be modified
significantly. If the cavity created by the stellar wind of
a massive star is small, e.g., with a radius
$\la 50$ pc, the supernova blast of the
star later on can catch up with the stellar wind and continue to drive
the high velocity region according to eq.~(\ref{Mv})
until radiative losses become important (McCray \& Kafatos 1987).
This is then an intermediate case between a pure stellar wind and a pure
SN shock, and the required $R_{\rm SN}$ will not deviate significantly
from eq.~(\ref{SNrate}) either.

For single halo absorber models (Pettini et al. 1983),
$r_{\rm abs}\sim 70h^{-1}$kpc, $M_{\rm abs}\sim 10^{11}M_\odot$,
eq.~(\ref{SNrate}) gives
$R_{\rm SN}>0.015$ per year per 10$^{11}M_\odot$, yielding
1.5$\times 10^7$ SNe in 10$^9$ years (which is a reasonable time frame since
the detection of these clustered CIV absorption systems ranges
from $z\approx 1.2$ to $z\approx 3$ (SBS)).
Since each type II SN ejects about $0.1M_\odot$ iron and type I
SN ejects 6 times as much (Truran \& Burkert 1994), this rate translates into
at least $1.5\times 10^7M_\odot$ iron in a 10$^{11}M_\odot$ galaxy,
or [Fe/H]$>-2$, in 10$^9$ years.
For dwarf galaxy absorber models (York et al. 1986),
$r_{\rm abs}\approx 8$kpc, $M_{\rm abs}\sim 2\times 10^9M_\odot$,
eq.~(\ref{SNrate}) gives
$R_{\rm SN}>0.0025$ per year per 2$\times 10^9M_\odot$, or
$0.12$ per year per $10^{11}M_\odot$, which yields
at least $1.2\times 10^8M_\odot$ iron in a 10$^{11}M_\odot$ galaxy,
or [Fe/H]$>-1$, in 10$^9$ years.

In mini-absorber models (Sargent et al. 1979),
the above calculation only applies for mini-absorbers with
mass $M_{\rm abs}>10^4M_\odot$, since for absorbers comparable to or
less than $10^4M_\odot$, a single SN will disrupt the whole cloud and
populations of mini-absorbers will soon be destroyed.
For mini-absorbers with a larger mass, such as $10^5M_\odot$
and a radius of 100 pc, eq.~(\ref{SNrate}) yields
$R_{\rm SN}>10^{-5}$ per year per absorber, or
$R_{\rm SN}>5$ per year per 10$^{11}M_\odot$ if we consider that
$5\times 10^5$ such mini-absorbers are needed to match the
cross section of a single 70 kpc galactic halo. This SN rate
is unacceptably large since the resultant [Fe/H] after 10$^9$yr will
be larger than 0.6.

In other words, if the clustered CIV absorption lines in the range of
200km/sec$<\Delta V<$600km/sec are solely due to high velocity gases
driven by SN shocks (and/or strong stellar winds),
the resultant SN rate will be too large for
mini-absorber models such as that of Sargent et al. (1979), but
remain reasonable in terms of the resultant metal yields
in the single halo absorber model and the dwarf
galaxy absorber model.
The SN rates calculated in the single halo and dwarf galaxy absorber models
also lie close to the estimate of the current
SN rate, 0.005--0.1 per year per galaxy depending
on the morphology of the galaxy (Cappellaro et al. 1993). (The
consideration of different type of supernovae will not change our
conclusions significantly.)

Conversely, with a reasonable SN rate, high
velocity gases driven by SN shocks (and/or strong stellar winds)
play little role in the clustering
of absorption lines in the mini-absorber models, but may play a
significant role in the single halo or dwarf galaxy absorber models.
Especially when the absorbers had star-bursts
at the redshift the absorptions are detected, which are expected
in the early history of galaxies (Yanny 1990), their SN rates
would be significantly higher than the SN rate today. The role
of SN (and/or stellar wind) driven clouds in producing clustered QSO
metal absorption lines could then be even more relevant.

The number of the clustered absorption lines from these SN
(and/or stellar wind) driven clouds should scale as $d(t_{100{\rm km/sec}}
\langle A\rangle_{\Delta V\ge 200{\rm km/sec}})/dv$, or $(\Delta V)^{-4}$,
which is much steeper than that expected from a galaxy-galaxy
correlation function, eq.~(\ref{correl}). According to this dependence
law, given the total number of $\sim 20$ clustered absorption pairs
with $\Delta V>200$ km/sec in SBS (1988), the clustered absorption
pairs should fall below 1 at $\Delta V\ga 500$--600 km/sec
, which coincides with the observation
of SBS (1988) that the clustering of
absorption lines ends at $\Delta V\approx 600$ km/sec.
Although the low statistics of available data and the over-simplification of
our present calculations
do not allow a realistic comparison at the moment, further
data with better statistics can potentially test
the predicted steep dependence of the number of clustered metal
absorption lines on the splitting velocity, and thereby test
the role of SN (and/or stellar wind)
driven clouds in producing the clustered QSO metal absorption lines.

Our calculations have been based on a spherical symmetry of the absorber and
the high velocity region. Although for SNe or stellar winds,
a spherical geometry or a nearly spherical geometry is most natural,
a geometry severely deviates from a spherical symmetry (e.g., sheet-like) for
these high velocity regions may not be impossible.
If so, their cross sections of intercepting a QSO line of sight may increase
significantly. However, in such a geometry, the velocity is most likely in
the direction of stretching of the structures, and the velocity
projected in the direction of large cross section is small.

Another concern is the uniform distribution of gases assumed in our
calculation. If massive stars in absorbers mostly
reside in underdensed region, the high velocity regions
their stellar winds or SN shocks generate will have
cross sections larger than that calculated from a uniform absorber.
Intuitively, however, massive stars should form more likely in
overdensed regions and gas rich environment.

\section{Summary}
We showed that while rich clusters of galaxies cannot explain the clustering
of CIV absorption lines with velocity  dispersions in between 200km/sec and
$\sim 600$km/sec, high velocity gases driven by SNe and/or strong
stellar winds could provide a viable solution to these clustered
absorptions, provided that the QSO metal-line absorbers are
single galactic halos or dwarf galaxies.
This provides an alternative to the galaxy-galaxy
correlation explanation to these clustered absorbers. Furthermore, the
expected number of absorption pairs due to high velocity gases driven
by SNe and/or strong stellar winds decreases
much faster with an increasing velocity split $\Delta V$ than
that expected from the galaxy-galaxy correlation, which provides
a testable feature as the statistics of clustered QSO metal absorption lines
improves. It is also possible that both of these two scenarios,
i.e., stellar activities and the galaxy-galaxy correlation, work at
the same time. Then the significance of the SNe (or strong stellar winds)
driven gases in providing the clustered absorptions depends on the SN
rate in absorbers. For
single galactic halo models, high velocity gases driven by SN shocks
and/or strong stellar winds become important if the SN rate
is $\ga$0.015 per year per 10$^{11}M_\odot$; for dwarf
galaxy absorber models, these gases become important if the SN rate is
$\ga$0.12 per year per 10$^{11}M_\odot$. There rates are
in or near the range of the estimated SN rate today. Therefore, if
absorbers had star-bursts at the absorption redshift
(between $z\sim 1$ and $z\sim 3$), their SN rates then would be significantly
higher than the SN rate today, and the role of high velocity gases driven
by SN shocks and/or strong stellar
winds in providing the clustered QSO metal-line absorptions will be important.

\section{Acknowledgement}
The author thanks Don York, David Schramm, James Truran
for helpful discussions. This work is supported by
NSERC grant at Queen's University, and by CITA National Fellowship at
Canadian Center for Theoretical Astrophysics.
\beginapjbib

\bibitem Abbott, D. C. 1982, ApJ, 263, 723
\bibitem Bahcall, J. N., \& Spitzer, L. 1969, ApJ, 156, L63
\bibitem Bahcall, N. A. 1988, Ann. Rev. Astr. Ap. 26, 631
\bibitem Lilje, P. B., \& Efstathiou, G. 1988, MNRAS, 231, 635
\bibitem Loveday, J., Peterson, B. A., Efstathiou, G., \& Maddox, S. J. 1992,
ApJ, 390, 338
\bibitem McCray, R. \& Kafatos, M. 1987, ApJ, 317, 190
\bibitem Peebles, P. J. E. 1993, Principles of Physical Cosmology, Princeton
University Press, page 471
\bibitem Pettini, M., Hunstead, R. W., Murdoch, H. S. \& Blades, J. C. 1983.,
ApJ, 273, 436
\bibitem Sargent, W. L. W., Young, P. J., Boksenberg, A. \& Turnshek, D. A.
1979, ApJ, 230, 49
\bibitem Sargent, W. L. W., Boksenberg, A. \& Steidel, C. C. 1988, ApJS,
68, 539 (SBS)
\bibitem Spitzer, L., Jr. 1978, Physical Processes in the Interstellar
Medium (New York: Wiley)
\bibitem Truran, J. W. \& Burkert, A. 1995, Phys. Rep., in press.
\bibitem Weaver, R., Castor, J., McCray, R., Shapiro, P. \&
Moore, R. 1977, ApJ, 218, 377
\bibitem Weyman, R., Carswell, R. F., \& Smith, M. G. 1981, Ann. Rev. Astr.
Ap. 19, 41
\bibitem Yanny, B. 1990, ApJ, 351, 396
\bibitem York, D. G., Green, R. F., Bechtold, J., \& Chaffee, F. H., Jr.
1984, ApJ, 280, L1
\bibitem York, D. G., Dopita M., Green, R. \& Bechtold, J. 1986, ApJ, 311, 610
\bibitem York, D. G. {\em et al.} 1991, MNRAS, 250, 24
\endapjbib
\newpage
\noindent{\bf Figure Captions:}
\bigskip

\noindent{Figure 1. The geometry of a quasar line of sight passing through
a cluster of galaxies.}
\bigskip

\noindent{Figure 2. The geometry of high velocity gases driven by a supernova
shock.}
\end{document}